
\input phyzzx

\REF\ch{
 R. Courant and D. Hilbert, Methods of Mathematical Physics, Vol.2
(Interscience, John Wiley and Sons, New York, 1962); see chapters 5 and 6.}
\refend

\REF\abs{ R. J. Adler, M. Bazin, and M. M. Schiffer, Introduction to General
Relativity, 2nd ed.  (McGraw Hill, New York, 1975); see chapter 4 for the
characteristics of the Maxwell equations and chapter 8 for those of the vacuum
Einstein equations; the Kerr solution and the black hole null surface are
disucssed in chapter 7.}
\refend

\REF\penrose{ R. Penrose, Gen. Rel. and Grav. {\bf 12}, no. 3, 225 (1980).}
\refend

\REF\zhs{ H. M. Zum Hagen and H.-J. Seifert, Gen. Rel. and Grav. {\bf 8},
no. 4, 259 (1977).}
\refend

\REF\kerr{ R. P. Kerr, Phys. Rev. Letters, {\bf 11}, 237 (1963); the Kerr
solution representing  a rotating black hole is built from a metric containing
 a null vector; see also ref. 2, chapter 7. }
\refend

\REF\hawking{ S. Hawking, Commun. Math. Phys. {\bf 43}, 199 (1975).}
\refend

\REF\bd{ N. D. Birrell and P. C. W. Davies, Quantum Fields in Curved Space
(Cambridge University Press, New York, 1982); see chapter 8. }
\refend

\REF\bmccpp{ S. J. Brodsky, G. McCartor, H. C. Pauli, and S. S. Pinsky,
Particle World {\bf 3}, no. 3, 109 (1993).}
\refend

\REF\countzm{O. C. Jacob, Phys. Lett. {\bf B347}, 101, 1995.}
\refend

\REF\ngc{ O. C. Jacob, Phys. Rev. {\bf D50}, 5289 (1994).}
\refend

\REF\pertre{ O. C. Jacob, Phys. Rev. {\bf D51}, 3017 (1995).}
\refend

\REF\iz{ C. Itzykson and J.-B. Zuber, Quantum Field Theory (McGraw Hill, New
York, 1980); see chap. 1.}
\refend
\REF\mp{ C. Morosi and L. Pizzocchero, J. Math. Phys. {\bf 35}, 2397 (1994).}
\refend

\REF\sovoliev{ V. O. Sovoliev, J. Math. Phys. {\bf 34}, 5747 (1994).}
\refend

\REF\lv{ J. Losa and J. Vives, J. Math. Phys. {\bf 35}, 2856 (1994).}
\refend

\PHYSREV
\overfullrule0pt

 \newtoks\slashfraction
 \slashfraction={.13}
 \def\slash#1{\setbox0\hbox{$ #1 $}
 \setbox0\hbox to \the\slashfraction\wd0{\hss \box0}/\box0 }

  \def\Buildrel#1\under#2{\mathrel{\mathop{#2}\limits_{#1}}}

\def\lozenge{\boxit{\hbox to 1.5pt{%
             \vrule height 1pt width 0pt \hfill}}}

 \doublespace
 \pubnum{7025 \cr
 hep-th/9503nnn}
\date{October  1995}
 \pubtype{T}
 \titlepage
 \title{Null Surfaces, Initial Values and
 Evolution Operators for Scalar Fields
 \doeack}
 \author{ Ronald J. Adler\foot{Permanent Address:
\it{
Department of Physics and Astronomy,
 San Francisco State University,
 San Francisco, CA 94132}
}
and
 Ovid C. Jacob\foot{Permanent Address:
\it{
Department of Physics and Astronomy,
 Sonoma State University,
 Rohnert Park, CA 94928}
}
\foot{
jacob@unixhub.slac.stanford.edu
}
}
 \SLAC

\abstract
We analyze the initial value problem for scalar fields obeying
the Klein-Gordon  equation. The standard Cauchy initial value problem
for second order differential equation is to construct a solution function in
a neighborhood of space and time form values of the function and its time
derivative on a selected initial value surface. On the characteristic surfaces
the time derivative of the solution function may be discontinuous, so the
standard Cauchy construction breaks down. For the Klein-Gordon equation the
characteristic surfaces are null surfaces. An alternative version of the
initial data needed differs from that of the standard Cauchy problem, and in
the case we discuss here the values of the function on an intersecting pair
of null surfaces comprise the necessary initial value data. We also present
an expression for the construction of a solution from null surface data; two
analogues of the quantum mechanical Hamiltonian operator determine the
evolution of the system.

\vfill
\centerline{PACS: 11.10 Cd, 11.10Ef, 11.30Er, 11.15-q}
\centerline{Submitted to J. Math Phys. }
\chapter{Introduction}
The standard Cauchy problem is to use a partial differential equation
and the values of the solution function and some of its derivatives
given on an initial value surface to determine the values of the function
in a neighborhood of the surface
[\ch]
The derivatives needed for an $n^{th}$ order partial differential equation
in time are the first $n-1$ time derivatives, and in applications to physics,
a surface often used is three-space at some initial time. Solution of the
standard Cauchy problem proceeds by showing how initial  value data and the
differential equation determine all the time derivatives of the function on the
initial surface and thus allow a power series development of the function
for future times.

If the solution function has a discontinuity in its $n-1$ time derivative
across some special surface the the standard Cauchy problem cannot be solved
using that surface for the initial value  data. Such special surfaces, the
characteristic surfaces, or simply characteristics, have been studied
extensively for many of the differential equations of physics. The physical
significance of characteristics is that the solution or a derivative on one
side may differ discontinuously from that on the  other side, so the
characteristic can represent a physical discontinuity  interpretable as a
wave front or shock wave. For example, the characteristics of the Maxwell
equations in free space are the wave fronts of light, that is null surfaces
[\abs].
 The characteristics of the Einstein equations in free space are also
null surfaces, so that gravity may be interpreted to propagate at the
velocity of light [\abs],
[\penrose], [\zhs].

 The characteristics of the Klein-Gordon equation for free particles
in flat space are easily shown to be null surfaces, as are the characteristics
of a large class of interacting versions of the Klein-Gordon equation.
Thus the derivative discontinuities of the solution functions propagate at
the speed of light, independently of the mass parameter in the equations.
Note in particular that the discontinuities do not propagate at either
the phase or group velocity of the wave solutions.

For many years, the use of null coordinates - coordinates lined up with
the null surfaces - has been common in general relativity. For example,
many problems involving black holes are conveniently studied with null
coordinates since black hole surfaces are time independent null surfaces
[\hawking], [\bd].
Recently interest has developed in using null coordinates in the
study of quantum field theories such as quantum electrodynamics and
quantum chromodynamics this is often referred to as light-cone quantization,
although the planar surfaces $x = \pm t$ are usually used
[\bmccpp].
In quantum field theories the behavior of the field operators is related
to the behavior of the classical (c-number) fields. For  example, the
propagator may be interpreted as either the Green's function of the classical
field or as the vacuum expectation value of the time-ordered field
operators
[\ngc].
It is thus possible that the discussion of this paper
may have implications for studies of black hole  physics in curved space or
quantum field  theories in flat space. Regarding the latter topic, the
transition from classical Green's functions to quantum vacuum expectation
values is beset with subtleties stemming from the possible appearance of
anomalies upon quantization
[\countzm].

  We have studied the Klein-Gordon equation and its characteristics, in
particular how the values of the solution function on characteristics determine
the values of the fields in a neighborhood of space and time. Operators
that determine how the solutions develop in time are usually called
evolution operators, and we use here the same term to include similar operators
associated with null coordinates.

We first briefly review the standard Cauchy problem for the Klein-Gordon
equation, and give an operator expression for developing the solution from
a constant time surface using as initial value data the function and its
first time derivative. We then obtain the characteristics and solve the
analogous problem for the planar null surfaces $x = \pm t$. Initial data
for this situation consists of the value of the function on the {\bf pair}
of null surfaces, and does {\bf not} include derivatives. We give a simple
evolution expression in terms of null coordinates, which is a direct analogue
to the usual expression of quantum theory, except that two analogues of
the Hamiltonian operator determine the evolution solution
[\pertre], [\iz], [\mp], [\sovoliev], [\lv].
\chapter{Development of a Scalar Function From a Constant Time Surface}
The standard form of the Cauchy problem for the Klein Gordon equation is to
determine how a solution may be developed from the appropriate initial value
data
on a constant surface, which we take to be $t = 0$. (See figure 1.)
The Klein Gordon equation for the scalar function $\phi (\vec {x}, t)$ is
$$ \phi_{|t|t}(\vec {x},t) - \bigtriangledown ^2 \phi(\vec{x},t) + m^2
\phi (\vec {x},t) = 0 \eqn\kget$$
(We use the units in which $\hbar=c=1$. The slash notation indicates
differentiation
with respect to the indicated variable, in this case time t.)

To solve the Cauchy problem we use the Klein Gordon equation and the value of
the
solution function and its first time derivative on the surface $t = 0$ to
compute
the value of the solution at future times. For this we use a Taylor series
expansion in t, which we write as
$$\phi(\vec {x},t) = \phi(\vec{x},t)_{t=0} +\phi_{|t}(\vec{x},t)_{t=0} t+
\phi_{|t|t}(\vec{x},t)_{t=0} {t^2\over2!}+... \eqn\taylors$$
We specify the solution and its first time derivative at $t = 0$ as
$$\phi(\vec {x},t)_{t=0} = h(\vec{x}) \quad \phi_{|t}(\vec{x},t)_{t=0} =
k(\vec{x})
\eqn\inidat     $$

Then the Klein Gordon equation allows us to calculate all the remaining time
derivatives on the $t = 0$ surface needed in the series expansion; the second
derivative is

$$\phi_{|t|t}(\vec {x},t)_{t=0} = [\bigtriangledown^2 \phi(\vec{x},t)
-m^2 \phi(\vec {x},t)]_{t=0} = [\bigtriangledown^2 - m^2]h(\vec{x})
\eqn\fitt $$

Higher derivatives are obtained in the same way, as

$$\phi_{|2nt}(\vec {x},t)_{t=0} = [\bigtriangledown^2 - m^2]^n h(\vec{x})
\equiv (-1)^n E^{2n} h(\vec{x}) \quad
(even\quad  or\quad 2n\quad derivative) $$

$$\phi_{|(2n+1)t}(\vec {x},t)_{t=0} = [\bigtriangledown^2 - m^2]^n k(\vec{x})
\equiv (-1)^n E^{2n} k(\vec{x}) \quad   (odd\quad or\quad 2n+1\quad
derivative)\eqn\hiderivs $$

Here $E^2$ is the Klein Gordon energy squared operator, $E^2 = m^2 -
\bigtriangledown^2 $.
Now the series \taylors may be written as

$$\phi(\vec {x},t) = \sum_{0}^{\infty}(-1)^n{(Et)^{2n}\over(2n)!}h(\vec{x}) +
\sum_{0}^{\infty}(-1)^n{(Et)^{2n+1}\over(2n+1)!}{k(\vec{x})\over E} \eqn\sols
$$

This may be formally summed to yield a concise result,

$$\phi(\vec {x},t) = cos(Et)h(\vec{x}) + {sin(Et)\over E}k(\vec{x}) \eqn\kgsol
$$

which displays the solution in terms of evolution operators acting on the
initial value functions $h(\vec {x})$ and $k(\vec{x})$ given on the initial $t
= 0$ surface.

The formal result \kgsol is directly usable if $h(\vec {x})$ and
$k(\vec {x})$ are expressed as Fourier transforms; the operator E may then be
replaced by the value $\pm\sqrt{m^2+p^2}$, for the p-th Fourier component. For
the p-th component the appropriate $h(\vec{x})$ is $e^{i\vec{p}.
\vec {x}}$, and the appropriate $k(\vec{x})$ is $-iEe^{i\vec{p}.
\vec {x}}$. Substituting these into \kgsol one obtains the correct function
$e^{-iEt +i\vec {p}.\vec{x} }$  for the p-th component. It is thus clear that
\kgsol indeed represents the general solution of the Klein Gordon equation.

So far we have limited our discussion to the free Klein Gordon equation \kget.
However if there is an additional interaction term in the equation which is a
function of the scalar field and time and position,
$F(\phi,\vec {x},t)$, then it is evident that the same considerations allow one
to determine higher time derivatives just as in
\fitt and \hiderivs, so that a power series solution will also exist in this
more general case.

\chapter{Characteristics of the Klein Gordon Equation}

Surfaces other than $t = 0$ may be used for the series development of the Klein
Gordon solutions; indeed only a limited class of surfaces does not allow such a
series development - the characteristic surfaces. On these surfaces the
standard Cauchy problem breaks down.

Let us attempt to develop a solution of the Klein Gordon equation from a
surface S, given by $t=T(\vec {x})$.
(See figure 1.) We suppose that the solution and its first time derivative are
given on S by initial value functions

$$\phi(\vec {x},T(\vec{x})) = \phi(\vec{x},t)_{t=T} = h(\vec{x}), \quad
\phi_{|t}(\vec{x},t)_{t=T} = k(\vec{x}) \eqn\initdatchar$$

 For brevity we write $t=T$ to denote $t=T(\vec {x})$. We then represent
the solution as a Taylor series in the form

$$\phi(\vec {x},t) = \phi(\vec{x},t)_{t=T} +[t-T(\vec{x})]
\phi_{|t}(\vec {x},t)_{t=T}+
{[t-T(\vec{x})]^2\over2}\phi_{|t|t}(\vec{x},t)_{t=T}+... \eqn\taylorschar$$

A critical step is to find the second time derivative in this expansion. To do
this we first relate derivatives of the initial data functions to derivatives
of the solution function $\phi$ on S. From \initdatchar we obtain immediately

$$\bigtriangledown h(\vec {x}) = \bigtriangledown \phi(\vec{x},t)_{t=T} +
\phi_{|t}(\vec{x},t)_{t=T}\bigtriangledown T(\vec{x})$$
$$\bigtriangledown k(\vec {x}) = \bigtriangledown \phi_{|t}(\vec{x},t)_{t=T} +
\phi_{|t|t}(\vec{x},t)_{t=T}\bigtriangledown T(\vec{x})
\eqn\charderiv $$

where the gradient on $\phi$ represents the partial with respect to the 3
spatial variables. We differentiate this again and rearrange terms using
\charderiv to obtain

$$\bigtriangledown^2 h(\vec {x}) =\bigtriangledown^2 \phi(\vec{x},t)_{t=T}+
\phi_{|t}(\vec{x},t)_{t=T}\bigtriangledown^2 T(\vec{x}) -
\phi_{|t|t}(\vec{x},t)_{t=T}\bigtriangledown T(\vec{x})^2 + 2\bigtriangledown
k(\vec{x}) . \bigtriangledown T(\vec{x}) \eqn\secondderiv $$

Using \secondderiv, the Klein Gordon equation, and the definitions \initdatchar
for the initial value functions we get the desired equation for the second time
derivative of $\phi$ on S,

$$(1-\bigtriangledown T(\vec {x})^2)\phi_{|t|t}(\vec{x},t)_{t=T}
=(\bigtriangledown^2 -m^2)h(\vec {x})-  k(\vec{x})\bigtriangledown^2 T(\vec{x})
-  2\bigtriangledown k(\vec{x}) . \bigtriangledown T(\vec{x}) \eqn\charcondeq
$$

If the square of the gradient of T is not equal to 1 we may solve this as in
section 2, and develop a power series solution in a similar manner. Indeed the
previous case corresponds to the function T having zero gradient and Laplacian.
However If the gradient squared is equal to 1 then the equation cannot be
solved for the second time derivative and the surface is a characteristic. Thus
the equation for the characteristics of the Klein Gordon equation is

$$ \bigtriangledown T(\vec {x})^2 = 1 \eqn\chareq $$

It is easy to see that this is the equation of a null surface - that is the
normal to S is a null vector; to do this we write the equation of S as

$$ t - T(\vec {x}) = G(x^{\alpha}) \eqn\chardef $$

The normal to S is the 4-gradient of G, or

$$n_{\alpha} = G_{|\alpha} = (1, -\bigtriangledown T(\vec {x}) )
\eqn\nsdef$$

Thus n is a null vector by \chareq, and S is a null surface (figure 2).

In the next sections we will study the particular null surfaces for T taken as
$\pm x$, or in terms of the commonly used null coordinates u and v (figure 3),

$$u = t-x = 0 \quad  	u\quad characteristic\quad or\quad null\quad surface $$
$$v = t+x = 0 \quad 	v\quad characteristic\quad or\quad null\quad surface
   \eqn\uvdefs$$

It is interesting that the characteristics are independent of the mass m. This
is an example of a more general fact, that if there is an interaction term of
the form $F(\phi,\vec {x},t)$
in the Klein Gordon equation then an equation analogous to \charcondeq is
obtained, with the same left hand side. Thus the equation for the
characteristics is independent of the extra term and the characteristics are
again null surfaces.

Equation \charcondeq gives the equation of the characteristics, but it also
tells us that the functions $h$ and $k$ on the characteristics are constrained
if the solution is regular, since the right side of \charcondeq is zero. Such
constraints are typical of characteristics, and in this case there is a simple
physical interpretation. To see this consider the characteristic $t = x$, and
let $\phi$ be a plane wave of the form $e^{-iEt + kx}$.
Then setting the right side of \charcondeq to zero we see that $E^2 = k^2 +
m^2$,
the usual energy-momentum or mass shell relation.

\chapter{ Construction of a Solution Function From Null Surface Data}

We now show how to construct solutions to the Klein Gordon using initial data
on null surfaces; the pair of null surfaces $u = t - x = 0 $ and
$v = x + t = 0 $ will be used. In terms of the null coordinates u and v the
Klein Gordon equation is

$$4\phi_{|u|v}(u,v,x_\perp) +\bigtriangledown_\perp ^2 \phi(u,v,x_\perp) +
m^2 \phi_{|u|v}(u,v,x_\perp) = 0 \eqn\kqns$$

Here the perpendicular symbol $\bigtriangledown _\perp$ refers to the y, z
subspace; in the remainder of this paper we will suppress that subspace
dependence and write the Klein Gordon equation in the form

$$\phi_{|u|v}(u,v) =-\mu^2\phi(u,v), \quad \mu^2 = {m^2\over4} \eqn\kgnstd$$

For the case of $\mu = 0$ this is the wave equation and has particularly simple
solutions - any differentiable function of u or v. We accordingly expand about
this special massless case by setting up a solution as a power series in
$\mu^2$, written explicitly as

$$\phi(u,v) = \phi^{(0)}(u,v) + \mu^2 \phi^{(1)}(u,v) + \mu^4
\phi^{(2)}(u,v)+... \eqn\museriessol$$

We substitute this into \kgnstd and equate powers of $\mu^2$ to obtain a
perturbative solution. The zeroth order equation and solution are

$$\phi^{(0)}_{|u|v} = 0, \quad \phi^{(0)}(u,v) = f(u) + g(v) \eqn\zeroordsol$$

where $f$ and $g$ are differentiable functions.  The first order equation is

$$\phi^{(1)}_{|u|v} = - \phi^{(0)}(u,v) = -[f(u) + g(v)] \eqn\firstordeq$$

This is an inhomogeneous linear equation, so the solution is composed of a
homogeneous part and a particular part. But the homogeneous part is of the same
form as the solution of the zeroth order equation \zeroordsol, and can be
absorbed into that solution . Thus we need only obtain a particular solution to
\firstordeq, which is elementary. We write it in the form

$$\phi^{(1)}(u,v) = -[\int_{u_{0}}^u f(u')du'](v-v_0) -[\int_{v_{0}}^v
g(v')dv'](u-u_0) \eqn\firstordsol $$

The parameters $u_0$ and $v_0$ are arbitrary constants and correspond to
arbitrariness in the functions $f$ and $g$.

Higher orders are handled in the same way. The second order solution is

$$\phi^{(2)}(u,v) = -[\int_{u_{0}}^u du'\int_{u_{0}}^{u'}
f(u'')du'']{(v-v_0)^2\over2} -[\int_{v_{0}}^v dv' \int_{v_{0}}^{v'}
g(v'')dv'']{(u-u_0)^2\over2} \eqn\secordsol $$

The series solution can be written as

$$\phi(u,v) = \sum_{n=0}^{\infty}{[-\mu^2(v-v_0)\Gamma_u]^n\over n!} f(u) +
            + \sum_{m=0}^{\infty}{[-\mu^2(u-u_0)\Gamma_v]^m\over m!} g(v)
\eqn\phitdsol $$

where the multiple integral operator $\Gamma^n_u$ on f is defined as

$$\Gamma^n _u f(u) = \int_{u_{0}}^u du' \int_{u_{0}}^{u'} du''
...\int_{u_{0}}^{u^{n-1}} du^n f(u^n)  \eqn\integoper $$

The operators $\Gamma^n_{u,v}$ are linear and the series \phitdsol may be
summed to give a remarkably concise result in terms of a pair of exponential
evolution operators,

$$\phi(u,v) = e^{-\mu^2(v-v_0)\Gamma_v} f(u)+e^{-\mu^2(u-u_0)\Gamma_u} g(v)
\eqn\cmpctsl$$

This shows explicitly how the solution is given in terms of the two generating
functions $f(u)$ and $g(v)$, right moving and left moving waves which are
related to initial value data as we will discuss below.

We illustrate the simplest case of using \cmpctsl and consider the evolution
operator acting on a constant $f(u) = 1$. The terms in the series \phitdsol are
trivially obtained by integration and give

$$\phi(u,v) = \sum_{n=0}^{\infty} {(-\mu^2 u v)^2\over(n!)^2} = F_c
(uv)\eqn\constsol $$

where we have taken $u_0 = v_0 = 0$ for convenience without loss of generality.
This is easily verified to be a solution of the Klein Gordon equation by
substitution; moreover we recognize the series as the Bessel function

$$F_c (uv) = J_0(2\mu\sqrt{uv}) \eqn\besselsol $$

This special solution will be useful in the discussion of exponential solutions
below.

Consider the values of $\phi$ on the null surfaces $u = u_0 = 0$ and $v = v_0 =
0$. We see that

$$\phi_0 (u,0) = f(u) + g(0), \quad \phi_0 (0,v) = f(0) + g(v) \eqn\phizero$$

Thus up to a constant the generating functions are the initial values on the
pair of null surfaces, representing right and left moving plane waves. The
value of either of the generating functions at $0$ may be chosen arbitrarily,
and we choose for convenience both equal to ${\phi_0\over2}$ at the
intersection of the surfaces, so that

$$f(u) = \phi_0(u,0) - {\phi_0(0,0)\over2}, \quad g(v) = \phi_0(0,v) -
{\phi_0(0,0)\over2} \eqn\fginit $$

We emphasize that the initial value data needed for the null surface
development consists of the function on a pair of null surfaces; it is not the
same as the data needed in the standard Cauchy problem, which is the function
and its first time derivative on an initial surface [\sovoliev], [\lv].

\chapter{ Plane Wave Solutions}

We next discuss plane wave solutions of the Klein Gordon equation, which form a
convenient set for the expansion of a general solution. We will display the
appropriate generating functions and verify that the evolution operators in
\cmpctsl give a correct solution. Plane wave solutions of energy E and momentum
k may be written in terms of $t$ and $x$ or $u$ and $v$ coordinates as

$$\phi(t,x) = e^{-iEt +ikx} = e^{-i\lambda u -i\tau v} \eqn\nspw $$

where the null momenta $\lambda$ and $\tau$ are related to $E$ and $k$ and the
mass parameter $\mu^2$  by

$$\lambda ={E+k\over2}, \tau = {E-k\over2}, \lambda\tau ={E^2 - k^2\over4} =
{m^2\over4} = \mu^2 \eqn\nsconst $$

For the plane waves the appropriate generating functions are thus

$$f(u) = e^{-i\lambda u} -{1\over2}, \quad g(v) = e^{-i\tau v} -{1\over2},
\quad \lambda\tau = \mu^2 \eqn\nspwinit  $$

We substitute these generating functions into \cmpctsl, abbreviating for
convenience $B = -i\lambda$, $C = -i\tau$, $\mu^2 = - BC$, and find

$$\phi(u,v) = (e^{BCv\Gamma_u}e^{Bu}+e^{BCu\Gamma_v}e^{Cv}) -
            (e^{BCv\Gamma_u}+e^{BCu\Gamma_v}){1\over2} \eqn\nspwbc$$

The last term in this we recognize as $F_c$ which we obtained as the series in
\besselsol. The first two terms are readily evaluated by expanding in double
series as

$$ (e^{BCv\Gamma_u}e^{Bu}+e^{BCu\Gamma_v}e^{Cv}) = \sum_{n=0,
j=0}^{\infty}{(BCv\Gamma_u)^n\over n!}{(Bu)^j\over j!} + \sum_{m=0,
k=0}^{\infty}{(BCu\Gamma_v)^m\over m!}{(Cv)^k\over k!} \eqn\powersersol $$

It is easy to obtain the operation of $\Gamma^n$ on powers of $u$, as

$$\Gamma_u ^n u^j = {u^{j+n}\over (j+1)(j+2) ... (j+n)} \eqn\gammapowers$$

We substitute this into \powersersol and rearrange summation indices to find

$$ (e^{BCv\Gamma_u}e^{Bu}+e^{BCu\Gamma_v}e^{Cv}) = \sum_{n\le
j}^{\infty}{(Cv)^n\over n!}{(Bu)^j\over j!} + \sum_{m\ge
k}^{\infty}{(Cv)^m\over m!}{(Bu)^k\over k!}  $$

$$  = \sum_{n=0, j=0}^{\infty}{(Cv)^n\over n!}{(Bu)^j\over j!}
+\sum_{n=0}^{\infty} {(-\mu^2 u v)^2\over(n!)^2} = e^{B u + C v} + F_c (uv)
\eqn\nsfirstterm  $$

Thus the $F_c$ part cancels and in summary of \nspwbc to \nsfirstterm we have

$$\phi(u,v) = e^{B u + C v} = e^{-i\lambda u -i\tau v} \eqn\nssolpw $$

which is the correct plane wave solution. It follows that the evolution
operator equation \cmpctsl will produce the correct solution in any case that
may be expanded as a Fourier transform.

\chapter{ Relation to Null Surface Quantization}

The form of the solution in \cmpctsl is suggestive of the quantum mechanical
time evolution operator $e^{-iHt}$. In fact the analogy can be made complete
and is the basis of "light cone" quantization schemes [\bmccpp], [\sovoliev],
[\lv].
In terms of the momenta $\lambda$ and $\tau$ associated with the null
coordinates $u$ and $v$ the energy-momentum relation is
$\lambda\tau = \mu^2 = {m^2\over4}$ from \nsconst. If we use the standard
operator replacements for these momenta

$$\Lambda \equiv i{\partial\over \partial u} \quad \Sigma \equiv
i{\partial\over \partial v} \eqn\ltoper $$

Then the the integral operators $\Gamma_{u,v}$ can be written symbolically in
terms of the momentum operators as follows,

$$\Gamma_u = {i\over\Lambda}, \quad \Gamma_v = {i\over\Sigma},
\eqn\lsinversoper$$

The expression \cmpctsl giving the solution can then be written as

$$ \phi(u,v) = e^{-iv{\mu^2\over\Lambda} }f(u) +e^{-iu{\mu^2\over\Sigma}
}g(v)$$

$$ = e^{-iv{\mu^2\over\Lambda} }[\phi_0(u,0)-{1\over2}]
+e^{-iu{\mu^2\over\Sigma} }[\phi_0(0,v)-{1\over2}] \eqn\nsopersol$$

The operators ${\mu^2\over\Lambda}$ and ${\mu^2\over\Sigma}$ are equal to
$\Sigma$ and $\Lambda$ by \nsconst and play the role of Hamiltonian operators
relative to $v$ and $u$; equation \nsopersol thus represents the operation of a
pair of Hamiltonians evolving a solution from the initial values on the pair of
null surfaces [\pertre], [\iz], [\mp], [\sovoliev], [\lv].

Note from the above discussion that initial value data on both null surfaces
are necessary for the evolution of a solution. From the discussion of the
exponential solutions it is also clear that the constant term in the generating
functions plays an important role in the evolution.

\chapter{Acknowledgements}
We  would like to
thank Prof. Stanley Brodsky       for his continuing support
as well as thank the SLAC Theory  Group for its hospitality.
\endpage
\refout
\end